\documentclass[11pt]{article}
\usepackage[utf8]{inputenc}
\usepackage{cite}
\usepackage{hyperref}
\usepackage{amsmath, amsthm}
\usepackage{nameref}
\usepackage[shortlabels]{enumitem}
\usepackage{todonotes}
\usepackage{graphicx}
\usepackage{subfig}
\usepackage{float}
\usepackage{natbib}
\usepackage{orcidlink}

\usepackage[hyphenbreaks]{breakurl}
\usepackage{xurl}
\hypersetup{
     colorlinks=true,
     breaklinks=true,
     linkcolor=blue,
     citecolor=magenta,
     filecolor=green,
     urlcolor=orange,
}

\title{Hybrid-DAOs: Enhancing Governance, Scalability, and Compliance in Decentralized Systems}
\author{Neil Shah,\orcidlink{0009-0009-6840-0346}}
\date{\today}

\begin{document}
    \maketitle
\begin{abstract}
Decentralized Autonomous Organizations (DAOs), based on block-chain systems such as Ethereum, are emerging governance protocols that enable decentralized community management without a central authority. For instance, UniswapDAO allows members to vote on policy changes for the Uniswap exchange. However, DAOs face challenges regarding scalability, governance, and compliance. \textit{Hybrid-DAOs}, which combine the decentralized nature of DAOs with traditional legal frameworks, provide solutions to these issues. This research explores various aspects of DAOs, including their voting mechanisms, which, while ensuring fairness, are susceptible to \textit{Sybil attacks}, where a user can create multiple accounts to exploit the system. Hybrid-DAOs offer robust solutions to these attacks, enabling more equitable voting methods. Moreover, decentralization can be understood through four properties: anonymity, transparency, accountability, and fairness, each with distinct implications for DAOs. Lastly, this work discusses legal challenges Hybrid-DAOs face and their promising applications across sectors such as nonprofit management, corporate governance, and startup funding. Overall, we argue that Hybrid-DAOs are the future of DAOs: the additional legal structure enhances the feasibility of many applications, and they offer innovative solutions to technical problems that plague DAOs.
\end{abstract}

\section{Introduction}

As blockchain technology continues to evolve, it has opened the door to new models of decentralized governance, with Decentralized Autonomous Organizations (DAOs) at the forefront. DAOs offer a way to distribute decision-making power across a wide network of participants, eliminating the need for centralized authorities. However, as DAOs gain prominence, challenges related to governance, scalability, and accountability have emerged, particularly in maintaining a balance between decentralization and legal compliance. This has led to the rise of Hybrid-DAOs, which blend decentralized principles with traditional legal frameworks to address these challenges.

The significance of Hybrid-DAOs lies in their potential to address critical shortcomings of fully decentralized governance. Hybrid-DAOs offer a way to bring stability, legal recognition, and adaptability to blockchain-based governance. Without these features, DAOs risk being limited in their capacity to scale effectively or to operate within existing legal systems, which restricts their applicability in broader contexts. For example, recent market data shows that DAO treasuries have experienced a downturn from \$37 billion to \$24.5 billion; while still substantial, it underscores the volatility and challenges facing fully decentralized organizations (see \cite{Total-2024-10-27}). Hybrid-DAOs present an opportunity to mitigate such risks by incorporating legal safeguards and compliance measures, making them more attractive to institutional investors and stakeholders.

This paper argues that Hybrid-DAOs have the potential to transform governance models across a range of sectors. By combining the transparency and participatory aspects of decentralized systems with the legal structure and accountability of centralized organizations, Hybrid-DAOs offer a more flexible and sustainable approach to decentralized governance. We delve into the practical applications of Hybrid-DAOs in areas such as nonprofit management, corporate governance, real estate, and participatory budgeting, providing a comprehensive analysis of their potential for integration and impact. In doing so, we aim to highlight the ways in which Hybrid-DAOs can address the limitations of fully decentralized systems and offer solutions for a more transparent and inclusive future. Furthermore, by leveraging financial data and real-world use cases, we aim to demonstrate the tangible benefits of Hybrid-DAOs, including increased investor confidence, reduced risk exposure, and enhanced operational efficiency.

\section{Background}
Blockchain technology, which utilizes techniques for ensuring the integrity of digital records through timestamping, first introduced by W. Scott Stornetta and Stuart Haber in 1991. Their work laid the foundation for a decentralized system that could verify the authenticity and chronological order of digital documents without relying on a central authority. By leveraging cryptography, they created a system where records could be securely linked together in a chain, with each new entry depending on the previous one, forming an immutable and tamper-proof ledger.

This concept of a secure, decentralized ledger was later expanded upon by an anonymous individual or group under the name Satoshi Nakamoto. In 2008, Nakamoto introduced the idea of using this technology as the backbone for a new form of digital currency—Bitcoin. (see \cite{bitcoin.pdf})

The core idea behind Bitcoin, and most other cryptocurrencies, is the creation of a decentralized, immutable ledger known as the blockchain. This technology records every transaction on the network, ensuring transparency and security. By enabling users to execute transactions directly on the network, blockchain eliminates the need for intermediaries like banks or payment processors, allowing individuals to exchange money directly through peer-to-peer transactions.

In a traditional transaction, such as swiping a card at a card reader, a transaction request is sent to the merchant's bank, which then contacts the card network (such as Visa) to verify the payment. The card network checks with the cardholder’s bank to ensure sufficient funds are available, approves or declines the transaction, and then relays this information back through the network to the merchant. However, with the advent of Bitcoin and other cryptocurrencies, new consensus protocols, or verification methods, have been introduced to ensure that all participants on the network agree on the transactions executed on the blockchain. These protocols, such as Proof of Work (PoW) and Proof of Stake (PoS), replace the need for centralized intermediaries by enabling decentralized verification and validation of transactions across the network. In other words, Visa is replaced by a large group of users, who collectively verify the integrity of transactions.

\subsection{Proof-of-Work vs. Proof-of-Stake}
Proof of Work (PoW), used by Bitcoin, operates by having computers, known as miners, solve complex cryptographic puzzles to validate transactions. When a miner successfully solves the puzzle, the solution (or block) is broadcasted to the entire network. Each node then updates its blockchain by appending the newly mined block, which can contain anywhere from a few hundred to over 2,000 individual transactions, depending on their size. While PoW was the first consensus method, it results in high energy consumption due to the intensive computational power required for mining.

Although not energy efficient, Proof of Work (PoW) has proven to be highly secure, as the validation of transactions and the addition of new blocks are distributed across a vast number of independent nodes. This decentralization ensures that no single entity has control over the blockchain. Combined with the computational difficulty inherent in PoW, this structure makes the Bitcoin network highly resistant to tampering and fraud. The immense computational power required to execute PoW has created a network often described as virtually unhackable under normal circumstances, provided the majority of the network participants remain honest.

Despite its strengths, PoW is not completely immune to attacks. One of the most notable vulnerabilities is the 51\% attack. In this scenario, if a single entity or group of miners gains control of more than 50\% of the network's total computational power, they could theoretically manipulate the blockchain. This could include double-spending, where the same cryptocurrency is spent more than once, blocking new transactions, or reversing recently confirmed transactions, thereby undermining the integrity of the network.

Another common attack against PoW is the Selfish Mining strategy. In this case, a miner or group of miners withholds newly mined blocks from the network, keeping them private instead of broadcasting them immediately. By doing so, the selfish miner secretly works on extending their private chain, aiming to eventually release a longer chain than the public one. When the private chain is finally broadcasted and accepted by the network, it becomes the dominant chain, allowing the selfish miner to collect rewards from both their private and public chains. This effectively increases their share of mining rewards at the expense of honest miners, disrupting the fairness and efficiency of the network (see \cite{Proof-2024-10-27}).

In recent years, Proof of Stake (PoS) has become the preferred consensus mechanism for many major cryptocurrencies. Compared to Proof of Work (PoW), PoS offers several advantages, including energy efficiency and scalability, which we discuss in detail below. Staking in PoS involves putting up collateral, similar to securing a loan with assets, to gain the right to propose the next block and earn transaction fees as a reward. This stake incentivizes validators to act honestly. If a validator attempts to validate fraudulent transactions, they risk having their stake slashed, meaning they could lose a portion or all of the cryptocurrency they have staked. Once a block is proposed, a committee of validators is chosen to attest to its validity. If the proposed block is found to be valid, it is added to the blockchain, and the proposing validator, along with the attesting validators, are rewarded. However, if a validator is found to have proposed an invalid block or acted maliciously, their staked coins can be slashed, ensuring the integrity of the network.

One of the strongest advantages of PoS is that it requires only a select few validators to operate at any given time, which contributes significantly to its energy efficiency. As mentioned earlier, this allows PoS networks to consume far less energy compared to PoW networks. For example, Ethereum, the second-largest blockchain, reduced its energy usage by 99.95\% when it switched from a PoW consensus mechanism to PoS in September 2022. Another key advantage of PoS is its scalability. The more predictable and efficient selection algorithm, combined with blocks being proposed at regular intervals, enables PoS to handle more transactions and validate blocks much faster than PoW. When Ethereum transitioned to PoS, the block time was reduced from 13-14 seconds to a consistent 12-second interval, speeding up the transaction processing time and ensuring a more stable network. Additionally, PoS lowers the technical barriers to entry for participating in the network. Unlike PoW, which requires significant investment in specialized hardware and incurs high energy costs for mining, PoS validators do not need as much computational power. However, this accessibility comes with its own drawbacks (see \cite{Proof-of-stake-2024-10-27})..

While similar concerns can be raised about PoW, PoS also faces the challenge of wealth-based influence within the network. Since validators are selected based on the amount of cryptocurrency they stake, those with larger holdings can exert more control over the network. This dynamic not only centralizes power among wealthier participants but also creates a higher financial barrier to entry for smaller holders, who may struggle to compete as validators due to the significant stake required (currently, 32 ETH or around 80,000 USD is required to become a validator).

\subsection{Blockchain Applications}
Much like traditional financial systems with many currencies, such as the US Dollar and the Swiss Franc, the blockchain space has expanded dramatically, giving rise to a wide array of cryptocurrencies. Coins like Ethereum have gained significant traction, each offering unique features. For instance, Ethereum introduced the concept of Smart Contracts -- self-executing contracts with terms directly embedded into code. These smart contracts have paved the way for decentralized applications (dApps), which operate without the need for centralized control. One prominent category of dApps are Decentralized Autonomous Organizations (DAOs). These organizations are governed by smart contracts and operate transparently on blockchain networks, enabling a more democratic decision-making process. DAOs have been applied in various contexts, from crowdfunding to collective ownership and cause funding, showcasing the potential of blockchain to revolutionize traditional organizational structures.

DAOs can be used for a variety of purposes, with some key examples including protocol changes, collections, and investments or grants. Protocol DAOs allow users who hold the DAO's native token (similar to how each country has its own currency) to vote on technology protocol changes, such as new features and upgrades. Collector DAOs are another common use case; for example, PleasrDAO, one of the largest funded DAOs, focuses on purchasing digital art. Similarly, FlamingoDAO invests in emerging artists and NFTs, showcasing another example of a Collector DAO (See \cite{DeFi-Beyond-Hipe}).

One notable example of a Collector DAO is ConstitutionDAO, which was formed with the goal of purchasing an original copy of the U.S. Constitution at auction. Although the bid was ultimately unsuccessful, ConstitutionDAO demonstrated the power of collective action and fundraising within a decentralized organization. One of the most significant applications of DAOs, however, is in investment and grant-giving. These DAOs, like Collector DAOs, raise funds and allow members to vote on different emerging Web3 projects to invest in or provide grants to, supporting the growth of the broader ecosystem (See \cite{delphidigitalWhatCollector}).

As we move forward, this paper will dive deeper into the intricacies of DAOs, examining their ethical, legal, and technological implications in shaping the future of society. Through this exploration, we aim to provide a comprehensive understanding of the current state and future potential of DAOs and blockchain technology, along with their diverse applications.

\section{Decentralized Autonomous Organizations and Voting Methods}
Decentralized Autonomous Organizations, or DAOs, are a new type of organization that operates without the need for centralized authority, such as an executive board or owner. Instead, DAOs are governed by smart contracts, which are self-executing contracts with terms prewritten into the code. These organizations rely on consensus and voting mechanisms to make decisions, allocate resources, and implement changes. Essentially, DAOs remove control from a concentrated leadership team and instead distribute power among the organization's members.

This distribution of power is the core idea behind the creation of decentralized organizations, as it gives the power back to the hands of the people. Since DAOs run on the blockchain, each one also has its own unique native token, which is a key component in voting. When a contract is proposed, members who hold the underlying token can vote on whether to pass or reject the proposal. The more tokens a member holds, the greater their influence on the vote is. There are many different voting mechanisms that balance the influence of members, which will be explored further in this section of the report (See \cite{citation-2024-10-28}.

\subsection{Examples of DAOs}
As covered in the prior section, DAOs can be started for many reasons, ranging from managing protocols for open-source projects to pooling investments and grants. UniswapDAO is a major Protocol DAO that governs the Uniswap Protocol, a decentralized exchange for trading different cryptocurrencies. Members who hold the \$UNI token have voting rights and can pass decisions guiding the future development of the platform, the fee structures, and its direction. In 2022, a Uniswap DAO member proposed activating the "fee switch," a governance decision that would allow token holders to receive a portion of the trading fees generated from transactions. Before moving to an on-chain vote, the DAO community discussed the proposal on decentralized forums to address concerns and refine it. One major argument against the change was that it could reduce rewards for liquidity providers, potentially decreasing liquidity on the platform. However, the proposal ultimately passed after the community agreed on a phased rollout to assess its real-world impact before full implementation.

Another example of a DAO is MakerDAO, which governs the Maker Protocol, enabling the creation of DAI, a stablecoin pegged to the value of the US dollar. Members of MakerDAO can vote on collateralization rates, risk mitigation strategies, and new developments within the ecosystem, such as adding new collateral types. This decision helped diversify the assets backing DAI and reduced the risks of relying too heavily on a single asset (like ETH) by allowing users to mint DAI tokens using additional collateral types such as USDC and WBTC.

However, not all DAOs work out as expected from the start. "The DAO," launched in 2016 on the Ethereum blockchain, just one year after Ethereum was created, was a venture capital fund that allowed members to pool money and vote on projects to fund. At its peak, "The DAO" held approximately 15\% of all ETH in circulation, making it one of the largest DAOs of its time. However, a flaw in the smart contract code allowed an attacker to exploit it and siphon off over \$60 million by making recursive calls to withdraw funds without the code updating the user's balance until after processing the withdrawal. The attacker was able to repeatedly initiate new withdrawals before their balance could be updated.

In response, the Ethereum community faced a difficult decision: either allow the exploit to stand or implement a hard fork to return the stolen funds to investors. Many argued that executing a hard fork would violate the core principles of blockchain—specifically, that it should be immutable and resistant to any external interference, even in the case of a security flaw, as doing so could set a dangerous precedent. However, others contended that correcting the hack was crucial to maintaining trust and ensuring the future growth of the Ethereum ecosystem. After intense debate within the community—through forums, discussions, and voting—Ethereum eventually underwent a hard fork, resulting in two blockchains: Ethereum Classic (ETC), which maintained the original chain, and Ethereum (ETH), which reversed the attack and returned the funds. These series of events highlighted the challenges and risks of decentralized organizations, as well as the complexities of community governance in blockchain ecosystems.
\subsection{Voting Mechanisms in DAOs and Arrow's Impossibility Theorem}

A major component of DAOs is the ability for users to vote on changes in policies and development, such as ETC’s hard fork. While this may seem straightforward at first glance, the reality is far more complex due to the nuances involved in vote counting and the design of fair voting mechanisms. Kenneth Arrow’s Impossibility Theorem, also known as Arrow’s Paradox, outlines four criteria that must be fulfilled for an ideal voting system: no dictator, impartiality, reflecting the will of the group, and independence of irrelevant alternatives. While it is mathematically impossible to fully satisfy all these conditions at once, the goal of voting systems in DAOs is to come as close to Arrow’s criteria as possible, balancing fairness, decentralization, and efficiency (See \cite{daovotingpower}).

The outcome of these votes plays a critical role in shaping protocol changes and determining funding decisions.
\subsection{Token-Based Voting}
The most common and straightforward method for determining the outcome of a vote in DAOs is the token-based voting system. In this approach, the influence of a member’s vote is directly proportional to the number of tokens they hold; more tokens equate to greater voting power. While simple and easy to implement, this system carries the risk of centralizing power. Since voting power increases linearly with the number of tokens, a small group of “whales” (individuals or entities holding large quantities of tokens) can exert disproportionate influence on the outcome, potentially undermining the decentralized nature of the DAO. As a result, votes may become skewed in favor of those with the most resources, limiting the impact of smaller coin holders. In this system, the outcome is ultimately determined by the majority of token-weighted votes, which may not always reflect the broader will of the community.

A common additional layer added to token-based voting is the requirement for a minimum amount of participation for a vote to be considered valid, which is called Quorum Voting. This system prevents decisions from being made by a small group not representing the whole and encourages a greater voter turnout; however, there still is a risk of stalled decisions if the quorum is too high and manipulation by whales (See \cite{Most-2024-10-28}).

\subsection{Quadratic Voting}
Quadratic voting is a variation of token-based voting that alters the way a token's influence is applied in decision-making. Unlike token-based and quorum voting, where each token grants one vote (1:1 ratio), quadratic voting progressively reduces the influence of each additional token. In quadratic voting, the influence a member has in the vote is proportional to the square root of the number of tokens they commit. This system is designed to balance the influence between large and small token holders, preventing whales from having an outsized impact while still allowing participants with more tokens to have greater, but not overwhelming, influence.

Mathematically, if \( T \) represents the number of tokens a member commits, the voting power \( V \) they receive is calculated as:
\[
V = \sqrt{T}
\]

This means that the more tokens a member holds, the diminishing return on influence applies. For example:
\[
\text{If } T = 100, \quad V = \sqrt{100} = 10
\]
\[
\text{If } T = 10,000, \quad V = \sqrt{10,000} = 100
\]

In this case, even though one individual has 100 times more tokens, the quadratic voting system creates a more balanced distribution of influence by assigning them only 10 times more voting power.

However, while this system helps prevent major concentrations of power, it is still vulnerable to a significant flaw: Sybil attacks. In a Sybil attack, a single entity creates multiple fake identities or wallets to gain disproportionate influence in the vote. For example, if a whale voting for Option B splits their 10,000 tokens across 100 wallets, each holding 100 tokens, the distribution of votes would be as follows:

1. Without Sybil Attack:
   \[
   T = 10,000, \quad V = \sqrt{10,000} = 100
   \]

2. With Sybil Attack (split across 100 wallets, each with 100 tokens):
   \[
   T = 100 \text{ tokens per wallet} \Rightarrow V = \sqrt{100} = 10 \text{ votes per wallet}
   \]
   \[
   \text{Total Votes} = 100 \text{ wallets} \times 10 \text{ votes} = 1,000
   \]

In this example, by splitting their tokens across multiple wallets, the whale effectively gains 10 times the voting influence, demonstrating the potential for manipulation within quadratic voting systems (See \cite{qv-2024-10-28}).

Later, we discuss solutions to mitigate such issues using identity verification mechanisms within a Hybrid-DAO model to reduce the risk of Sybil attacks.

\subsection{Conviction Voting}
Unlike Quadratic and Token/Quorum Voting, where the number of coins directly relates to voting influence with no other factors, Conviction Voting is based on the aggregated preference of the community. This means that the longer you keep your tokens on a certain choice, the greater the power of your vote becomes for that choice. However, if you change your decision to another option at any point, your voting power would reset. This mechanism makes Conviction Voting immune to last-minute attempts to ambush the vote or any sudden surge in interest from individuals.

The mathematical system behind Conviction Voting is relatively simple, and while the value of the decay can differ between implementations, the overall essence remains the same:
\[
\text{Conviction} = \text{tokens} \times \left( 1 - e^{-\text{time decay}} \right)
\]
This encourages voters not to change their vote, as doing so would reset the power they have accumulated. The decay value also plays an important role in preventing tokens from building full influence too quickly, instead spreading it across a longer time frame.

However, this system is not without flaws. Like Token and Quorum Voting, it can indirectly favor the wealthy and reduce the power of smaller holders. Additionally, Conviction Voting can result in entrenched interests, as individuals may vote for one option and, even if they later reevaluate and find another option more favorable, they may hesitate to switch their vote due to the loss of accumulated influence (See \cite{Most-2024-10-28}).

Now that we’ve explored various examples of DAOs and their operations, we’ll dive deeper into the motivations driving their creation. Specifically, we will conduct a detailed analysis of decentralization: what it entails, when it provides value, and how DAOs can effectively implement and benefit from it.

\section{Decentralization}

Decentralization refers to the distribution of power, decision-making, and operational functions across a wide network of independent participants rather than a single central authority. The most popular example of this is the Bitcoin Network, a peer-to-peer virtual currency. In decentralized systems, no single entity has complete control, which can lead to greater transparency, security, and resilience. This principle underpins the structure of DAOs, allowing them to operate autonomously and democratize decision-making. Here, we analyze the key virtues of decentralization, including fairness, accountability, transparency, and the often misunderstood concept of anonymity.

\subsection{Anonymity}
A common misconception is that decentralization inherently implies anonymity, largely because blockchain technology, on which many decentralized systems are built, often promotes pseudonymity through public keys and wallet addresses. This perception arises from blockchain's early association with privacy-focused applications like Bitcoin, where participants could transact without revealing their identities. However, decentralization and anonymity are distinct concepts. While blockchain systems can allow anonymity, it is not a requirement for decentralization to function.

Anonymity is seen as a core assumption in many early blockchain use cases due to the focus on privacy. However, this assumption doesn't always hold true, especially in systems like DAOs, where transparency and accountability are crucial. In contexts involving significant financial transactions or legal contracts, maintaining full anonymity can undermine trust and hinder accountability. In such cases, decentralization can coexist with identifiable participants, balancing collective governance and legal responsibility while still distributing decision-making power across the network.

\subsection{Transparency}
Transparency is one of the foundational virtues of decentralization. In decentralized systems, every action or decision is recorded on a publicly accessible ledger, such as a blockchain, where participants can verify the accuracy and legitimacy of transactions or proposals. This openness ensures that no single entity can manipulate or obscure information, as it is shared across the entire network. In the context of DAOs, transparency allows members to see how decisions are made, who voted for what, and how resources are allocated. This fosters trust among participants, as everyone has equal access to information, minimizing the potential for corruption or hidden agendas.

\subsection{Accountability}
Decentralization promotes both transparency and accountability by distributing power across many independent participants, unlike traditional organizations, where decision-makers can often avoid scrutiny. Transparency in decentralized systems like DAOs stems from open decision-making processes and the use of an immutable ledger that records all actions. However, accountability doesn't always require full transparency. Even without complete visibility, smart contracts can enforce accountability by automatically penalizing bad behavior through mechanisms like slashing. This ability to implement consequences without needing total transparency is a key distinction, ensuring participants are held responsible without needing to reveal their identity.

\subsection{Fairness}
Fairness in decentralized systems stems from the principle of equal participation and distributed decision-making. Rather than allowing a small group of individuals or centralized authority to dominate decisions, decentralization ensures that power is spread across the network. In DAOs, fairness is often achieved through voting mechanisms that give all token holders a voice, proportionate to their stake in the system. This allows for more democratic governance, where decisions are made based on collective input rather than the influence of a few. Fairness also ties into the transparency and accountability of the system, ensuring that decisions are made openly and that all participants have the opportunity to contribute.

\subsection{When Decentralization is and isn't Needed}

\subsubsection{When Complete Decentralization is Useful}
Complete decentralization is beneficial in several cases:
\begin{itemize}
    \item \textbf{No single point of failure:} In a centralized system, a compromised leader or a central server crash could bring down the entire operation.
    \item \textbf{True anonymous voting:} Anonymity allows members to express their true beliefs without fear of social pressure or judgment.
    \item \textbf{Censorship resistance:} Especially in politically sensitive or authoritarian regions, decentralization allows the free flow of ideas without the risk of suppression by governments.
    \item \textbf{Global reach:} Operating without major legal structures, decentralization allows anyone, anywhere in the world, to participate and collaborate.
\end{itemize}

\subsubsection{When Complete Decentralization is Not Useful}
Complete decentralization has certain limitations:
\begin{itemize}
    \item \textbf{Overcomplication and governance issues:} Fully decentralized systems can face challenges like Sybil attacks, where one person can create multiple fake identities to manipulate voting.
    \item \textbf{Accountability concerns:} Like "The DAO", complete decentralization holds the risk of a lack of accountability.
    \item \textbf{False sense of decentralization:} In some DAOs, true decentralization may be more illusory than real. For example, there may be a small group with developmental power (push to main on GitHub) or a concentration of large token holders who can undermine the system.
\end{itemize}

\subsection{Hybrid-DAOs}

While complete decentralization comes with its own risks and benefits, a new model is emerging, referred to in this paper as Hybrid-DAOs. Hybrid-DAOs blend decentralized governance with legal frameworks, placing greater emphasis on fairness, accountability, and transparency while still allowing for some level of anonymity. However, in these organizations, the focus often shifts away from anonymity when necessary to ensure compliance and trust. This approach allows Hybrid-DAOs to balance decentralized principles with the practical need for legal oversight and ethical accountability.

The concept of "progressive decentralization" is becoming increasingly popular in the DAO space. This approach involves launching a project with a centralized structure and gradually decentralizing governance as the community grows and the protocol matures (See \cite{Progressive-2024-10-28}). This method allows for a more controlled and secure rollout of decentralized governance while still working towards the end goal of full decentralization. It also provides an opportunity to test and refine governance mechanisms before handing over complete control to the community, reducing the risk of failures.

A notable example of this approach is Compound. Initially, Compound, a decentralized finance (DeFi) protocol, was managed by a small, centralized team of developers. As the platform grew, they introduced the \$COMP governance token, which gave the community voting power over protocol upgrades, parameter adjustments, and the addition of new assets. Over time, Compound progressively transitioned more decision-making power to token holders, ensuring that the protocol's decentralization increased in tandem with its growth and maturity (See \cite{What-2024-10-28}).

Another popular choice among many Hybrid-DAOs is combining elements of decentralization with traditional legal structures. For example, a DAO might register as a legal entity in a jurisdiction favorable to blockchain technology while maintaining decentralized governance for decision-making. This approach allows the DAO to operate within the bounds of the law while still adhering to the principles of decentralization. Wyoming’s DAO legislation, which recognizes DAOs as legal entities, is an example of how jurisdictions are beginning to adapt to this new organizational model, potentially paving the way for broader acceptance and integration of DAOs into the global economy.

By leveraging blockchain technology and smart contracts, Hybrid-DAOs offer a more democratic and transparent alternative to traditional hierarchical organizations. As the ecosystem continues to evolve, these DAOs are likely to play an increasingly prominent role in the digital economy; however, a key component of voting mechanisms has to be explored further by the blockchain and DAO community (See \cite{Decentralized-2024-10-28}).

\subsection{Voting Methods in Hybrid-DAOs}

In Hybrid-DAOs, the voting process becomes more intricate, as it must balance regulatory compliance, legal accountability, and fairness. The challenge lies in creating a system that preserves the decentralized nature of decision-making while ensuring that the voting process is transparent, accountable, and equitable.

Hybrid-DAOs offer a potential solution to Sybil attacks by incorporating identity verification mechanisms alongside decentralized governance. Unlike traditional DAOs, which prioritize complete decentralization and anonymity, Hybrid-DAOs can combine decentralized principles with legal frameworks to address identity-related challenges without undermining fairness (See \cite{DIDs-2024-10-28}). By using identity verification platforms like ID.me, for example, Hybrid-DAOs can prevent users from creating multiple fake identities or wallets, ensuring that each participant is only allowed one verified wallet. This ensures that voting is fair, with equal weight distributed among participants, while mitigating the risk of manipulation from Sybil attacks.

The voting mechanisms discussed earlier in this paper—Token-Based Voting, Quorum Voting, Quadratic Voting, and Conviction Voting—are central to DAO governance and take on even greater importance in the context of Hybrid-DAOs. Each method helps strike a balance between accountability, transparency, and fairness, which are crucial when considering legal and regulatory safeguards. For example, Quadratic Voting can ensure that wealthy participants do not have disproportionate influence, promoting fairness in decision-making.

As Hybrid-DAOs make decisions in high-stakes sectors like healthcare or finance, ensuring that voting is not only secure but also fair is vital. This means creating systems where all participants, regardless of their stake, have a fair chance to influence the outcomes without being overshadowed by larger holders or bad actors. The voting process must remain transparent and protective against manipulation, all while upholding the principles of decentralized governance.

Furthermore, these voting systems are highly adaptable and offer far more depth than existing models, such as the electoral college. By integrating fairness into their design, Hybrid-DAOs are positioned as an attractive alternative to traditional governance systems, as they move beyond simple majority voting to embrace more sophisticated, equitable approaches to decision-making.

\section{Hybrid-DAOs Explored}
\subsection{Legal Structure of Hybrid-DAOs}
The legal structure of Hybrid-DAOs is evolving with the introduction of new DAO-specific laws, particularly in the United States. In 2021, Wyoming became the first state to recognize DAOs as legal entities by allowing them to register as limited liability companies (LLCs). This legislation provides DAOs with critical legal protections, primarily by limiting the liability of participants. By registering as LLCs, DAO members are shielded from personal liability, meaning their private assets are protected from claims against the organization (See \cite{law-2021-2024-10-28}). Other jurisdictions, such as Vermont and Switzerland, are exploring similar frameworks that grant DAOs legal legitimacy while maintaining decentralized governance. These regulations mark a significant step in reducing risk for participants and enabling DAOs to operate within a secure, regulated environment.

Despite these advancements, there is still significant work needed to fully establish a comprehensive legal structure for Hybrid-DAOs. One complication is the uncertainty around which jurisdiction's laws should apply when enforcing DAO-related issues. A key challenge is also the lack of uniformity across jurisdictions. While Wyoming has taken the lead, many other states and countries have yet to introduce similar legal frameworks, creating a patchwork of regulations. This inconsistency complicates taxation, compliance, and liability for Hybrid-DAOs operating globally. Legally, several issues need to be addressed, including the recognition of token-based governance, the enforceability of smart contracts, and the legal standing of decentralized decision-making processes. Additionally, questions around how to classify DAO tokens (as securities or otherwise) and how to handle disputes within decentralized systems remain unresolved. Without clear legislation to address these concerns, the wider adoption of Hybrid-DAOs could face significant hurdles.

\subsection{Reconciling Decentralized Governance with Legal Obligations}
Hybrid-DAOs often involve decision-making distributed across a network of participants, which may seem incompatible with existing requirements for board meetings, fiduciary duties, and other corporate responsibilities. Traditional corporate governance relies on hierarchical structures and well-defined roles, while Hybrid-DAOs use decentralized processes like token-based voting. Instead of a fundamental conflict, this may simply require creating analogs for traditional concepts—such as regular on-chain proposal discussions to mimic board meetings or establishing collective fiduciary responsibilities for participants. Future legal frameworks will need to address how Hybrid-DAOs can balance decentralization with accountability, ensuring these organizations operate within legal boundaries while upholding core principles of transparency, accountability, and fairness.

\subsection{Applications of Hybrid-DAOs}

As Hybrid-DAOs rapidly gain traction, the need to explore new applications for this innovative model of governance and decentralization will become increasingly critical. Currently, the implementation of Hybrid-DAOs is limited due to the lagging development of legal frameworks that can adequately accommodate this structure. We propose five main applications of Hybrid-DAOs.

\subsubsection{Non-Profits}
A Hybrid-DAO model applied to non-profits could greatly enhance both transparency and accountability by logging all donations onto a blockchain, creating a permanent and visible record of every transaction. Hybrid-DAOs offer a participatory model, where donors can vote on how funds are allocated to specific projects or causes, giving them more control and involvement in decision-making. Beyond tracking donations, the use of smart contracts ensures that funds are automatically allocated to projects once predetermined conditions are met. By utilizing Hybrid-DAOs, nonprofits can create an ecosystem where donations are tracked, visible to the public, and distributed in a way that reflects the collective decisions of the donor community.

\subsubsection{Corporate \& Customer Engagement}
Hybrid-DAO models offer companies a way to engage not just shareholders but also their customer base by giving them a direct say in the development and governance of the business. This approach promotes transparency and empowers a more engaged community to participate in the company’s development. For instance, tokens could act like "store cash," allowing customers to use tokens to vote on product features or company initiatives and redeem them for merchandise or services. By blending financial and participatory incentives, Hybrid-DAOs foster a community-driven model that sets companies apart in the marketplace, enhancing transparency, loyalty, and customer satisfaction.

\subsubsection{Startup Funding}
The Hybrid-DAO model can also be applied to startup funding, where the community can pool funds and collectively invest in early-stage companies. Unlike traditional venture capital or investment groups, this model democratizes access to early-stage investments that were previously limited to accredited investors. Hybrid-DAOs offer enhanced transparency and liquidity compared to traditional startup funding methods. Token holders have liquidity options, as they can sell their tokens on secondary markets, providing flexibility that is often unavailable in conventional startup investments, where investors typically have to wait for an exit event to see returns.

\subsubsection{Real Estate Groups}
Similar to funding Hybrid-DAOs, this model could also be applied to shared ownership in real estate. Hybrid-DAOs take the concept of fractional ownership further by granting property owners more direct control and governance. Token-based voting allows participants to influence key decisions such as property maintenance, rent policies, and tenant selection. Hybrid-DAOs remove barriers by allowing participants to invest smaller amounts through the purchase of tokens, enabling investors to earn dividends from rent payments and enjoy liquidity by selling their tokens on secondary markets.

\subsubsection{Participatory Budgeting}
Hybrid-DAOs have significant potential to revolutionize participatory budgeting by offering citizens a more direct way to influence how public funds are allocated. In this model, rather than elected officials or government agencies deciding on how to distribute the budget for community projects, citizens themselves can vote on funding allocations for local needs such as parks, schools, or infrastructure. By recording all votes and budget allocations on a blockchain, Hybrid-DAOs ensure accountability in public spending, offering transparency and accessibility in decision-making.

\section{Conclusion}

This paper has delved into the foundational aspects of blockchain technology, tracing its development from the work of Stornetta and Haber in 1991 to its adoption in Bitcoin by Satoshi Nakamoto. We explored the evolution of consensus mechanisms, including Proof of Work (PoW) and Proof of Stake (PoS), and their implications for security, scalability, and energy efficiency in decentralized networks. In addition, we discussed the rise of Decentralized Autonomous Organizations (DAOs) and their application across sectors such as investment, governance, and protocol management. Furthermore, the concept of Hybrid-DAOs was introduced as a solution to the challenges posed by complete decentralization, offering a model that balances decentralized governance with legal and regulatory safeguards.

Looking forward, several key areas for future research and integration have been identified. First, legal and regulatory frameworks need further development to accommodate the unique structures of Hybrid-DAOs. Specific attention should be given to the issues of accountability and liability, particularly in systems where participants may remain anonymous or semi-anonymous. Additionally, research should explore the integration of Hybrid-DAOs into traditional sectors like real estate, healthcare, and public governance. For example, using Hybrid-DAOs in participatory budgeting and nonprofit management presents opportunities to enhance transparency and community involvement. Expanding on these applications will require legal clarity and robust technological frameworks.

Finally, future research must also address the technical and governance aspects of Hybrid-DAOs, especially in improving voting mechanisms and identity verification to prevent manipulation and maintain fairness. Quadratic voting, conviction voting, and other models should be refined to ensure equitable participation while balancing the needs of smaller and larger stakeholders. The ongoing development of identity verification systems integrated with blockchain governance will be critical in mitigating Sybil attacks and ensuring secure, transparent decision-making. By focusing on these areas, Hybrid-DAOs have the potential to significantly transform both decentralized and traditional organizational structures.

\bibliographystyle{ACM-Reference-Format}
\bibliography{citeBib/refs}
\nocite{*}

\end{document}